\documentclass[useAMS,usenatbib]{mn2e}
 
\usepackage{amssymb,amsmath}
\usepackage{graphicx} 
\usepackage{appendix}

\usepackage{color,soul}
\soulregister\citealt7
\soulregister\citet7
\soulregister\citep7
\soulregister\ref7

\usepackage[T1]{fontenc} 
\usepackage{aecompl}

\begin{document}

\title[Radio weak lensing shear measurement II]{Radio Weak Lensing Shear Measurement in the Visibility Domain - II. Source Extraction}
\author[Rivi \& Miller]{M. Rivi$^{1}$\thanks{E-mail: m.rivi@ucl.ac.uk}, L. Miller$^{2}$\\
$^{1}$Department of Physics and Astronomy, University College London, Gower Street, London, WC1E 6BT, UK\\
$^{2}$Astrophysics, Department of Physics, University of Oxford, Keble Road, Oxford, OX1 3RH, UK}

\maketitle

\begin{abstract}
This paper extends the method introduced in \cite{Rivi16} to measure galaxy ellipticities in the visibility domain for radio weak lensing surveys. In that paper we focused on the development and testing of the method for the simple case of individual galaxies located at the phase centre, and proposed to extend it to the realistic case of many sources in the field of view by isolating visibilities of each source with a faceting technique.
In this second paper we present a detailed algorithm for source extraction in the visibility domain and show its effectiveness as a function of the source number density by running simulations of SKA1-MID observations in the band 950-1150~MHz and comparing original and measured values of galaxies' ellipticities. Shear measurements from a realistic population of $10^4$ galaxies randomly located in a field of view of 1~$\deg^2$ (i.e. the source density expected for the current radio weak lensing survey proposal with SKA1) are also performed. 
At SNR $\ge$ 10, the multiplicative bias is only a factor 1.5 worse than what found when analysing individual sources, and is still comparable to the bias values reported for similar measurement methods at optical wavelengths. The additive bias is unchanged from the case of individual sources, but it is significantly larger than typically found in optical surveys. This bias depends on the shape of the uv coverage and we suggest that a uv-plane weighting scheme to produce a more isotropic shape could reduce and control additive bias.  

\end{abstract}


\section{Introduction}
\label{sec:intro}
Cosmological or targeted surveys of weak gravitational lensing at radio wavelengths may have a relevant role in the next years, when the Square Kilometre Array (SKA)\footnote{https://www.skatelescope.org} radio telescope will start to operate, providing a density of detected galaxies sufficient for shear measurement and a resolution to reliably measure their shapes. They will also be able to probe to higher redshifts given the different galaxy redshift distributions compared to the optical band \citep{Brown15}. Although the galaxy number density in the radio band will be lower than in the optical, the possibility to observe deeper can make radio weak lensing surveys for cosmology measurements competitive with the corresponding  optical surveys, as shown in recent forecasts from SKA simulations \citep{Harrison16}. Moreover, radio observations have the advantage of a deterministic knowledge of the image-plane Point Spread Function (PSF), being the Fourier Transform of the uv coverage, and will provide unique approaches for mitigating intrinsic alignments, such as concurrent measurements of polarization \citep{BB11} and galaxy rotation velocities \citep{Blain02, Morales06}.  Being subject to different observational systematics, cross-correlation with optical observations of the same field will allow suppression of systematic errors on shear measurement from future large surveys \citep{Patel10, DB16, Camera17}. This is quite relevant for precision cosmology as these errors may become comparable to, and larger than, the statistical noise.

The precursor radio weak lensing survey SuperCLASS\footnote{http://www.e-merlin.ac.uk/legacy/projects/superclass.html} is already underway and will soon provide data that may be used to test new methods required for accurate galaxy shape measurement in the radio band. A natural approach for such methods is working in the visibility domain where the data originates and the noise is Gaussian, avoiding non-linear data manipulation of the imaging process. SKA simulations have already shown that current imaging methods produce images with structures in the residuals which may dominate the cosmological signal \citep{Patel15}. Also cross-correlation analysis using real data images shows that no evidence of correlation is found between the optical and radio intrinsic shape of the matched objects \citep{Patel10,Tunbridge16}. This result suggests the presence of systematics in the procedure adopted for turning the visibility data into images, although a significant percentage of AGN sources in the observed population may be another possible explanation, as well as the astrophysical scatter between optical and radio position angle due to the different emission mechanisms in the two bands.

Currently, cosmic shear in the radio band has been successfully detected only working in the visibility domain but obtaining sources position from the image  \citep{Chang04}. Galaxies' ellipticities from  the VLA FIRST survey \citep{Becker95} were measured by decomposing them into shapelets, an orthonormal basis of functions corresponding to perturbations around a circular Gaussian invariant under Fourier transform \citep{CR02}. Since the data size and the number of resolved sources ($\sim 20-30 \deg^{-2}$) of each pointing is quite small, a joint fitting of the shapelet coefficients was possible by solving normal equations. Such an approach, computationally convenient, becomes very challenging when dealing with the order of $10^4$ sources per square degree and a very large dataset per pointing (order of PetaBytes), as expected from SKA Phase~1 continuum surveys \citep{Brown15}. Moreover shapelets introduce a shear bias as they cannot accurately model steep brightness profiles and highly elliptical galaxy shapes \citep{Melchior10}.  

In the companion paper (\citealt{Rivi16}, hereafter Paper~I) we presented \textit{RadioLens}fit, an alternative method working in the visibility domain where model fitting is performed on a single source at a time using an exponential profile as model for the galaxy. It is an adaptation of the optical Bayesian \textit{lens}fit method~\citep{Miller13} to radio data, where model visibilities are defined analytically and the likelihood is marginalised over uninteresting parameters. The method was tested in the simple case of individual galaxy visibilities simulated adopting the SKA1 uv coverage described in Section~\ref{sec:ska}, and the shear noise bias \citep{RKABR12, MV12} estimated as a function of the signal-to-noise ratio (SNR). Results  compared with requirements \citep{Brown15} for the proposed SKA1 radio weak lensing survey \citep{SKA1b} showed that the multiplicative shear bias may need calibration corrections similar to those for optical surveys, while the additive bias have to be controlled by an isotropic sampling of the visibility plane.  

In this paper we extend this work implementing the method for isolating source visibilities from realistic data, i.e. when many galaxies are in the field of view. We estimate its effectiveness in terms of ellipticity fitting and shear measurement by running SKA1-MID simulations as we did in the previous paper. We finally investigate the effect of the shape of the uv coverage on the additive shear bias. This paper is organised as follows. In Section~2 we summarise \textit{RadioLens}fit and present the extraction algorithm. In Section~3 details of SKA1 simulations are provided, while in Sections~4 and 5 results for galaxy ellipticity and shear measurements are presented respectively. Finally we discuss the shear additive noise bias in Section~6.

\section{Overview of RadioLensfit}
\label{sec:radiolensfit}
\textit{RadioLens}fit is a method for measuring radio galaxy ellipticities in the visibility domain. The idea is to adapt the approach used in the optical case to radio data, i.e. extracting from visibilities and model fitting a single source at a time.  Source extraction is difficult in the Fourier domain because signals from all sources in the primary beam are mixed altogether in the visibilities and sources are no longer localised. 
For this reason a joint analysis with the image domain may be needed: it allows us to identify sources and measure their position and flux with sufficient accuracy. With such information we can also compute a model of the observed sky and use it to approximate the signal from the other galaxies that must be removed  when extracting each source. The extraction is completed using a faceting technique that phase shifts the phase centre to the source position and further reduces its signal contamination by averaging visibilities in a coarse grid. Finally the model fitting can be performed as for the simple case of a single galaxy in the primary beam located at the phase centre as summarised in Section~\ref{sec:fitting}.
This way we can largely reduce the computational time when a huge number of sources are in the field of view (as for SKA) instead of trying a challenging joint fitting of all sources. A detailed algorithm for the fitting of many sources in the primary beam is presented in Section~\ref{sec:extraction}.

\subsection{Galaxy ellipticity fitting}
\label{sec:fitting}
In Paper~I we introduced this method as an adaptation of \textit{len}fit \citep{Miller13} by performing the chi-square fitting of single source visibilities. They are evaluated at the uv points, that are defined by the baselines formed between two antennae projected on the plane orthogonal to the pointing direction.
Model visibilities of a star-forming (SF) galaxy are computed analytically as the Fourier transform of the exponential brightness profile (S\'{e}rsic index $n=1$):
\begin{equation}
\label{model}
V(u,v) =  \Big( \frac{\lambda_\textrm{ref}}{\lambda}\Big)^\beta \frac{S_{\lambda_\textrm{ref}}\mathrm{e}^{2\pi \mathrm{i} (u l + v m)}}{\big(1+4\pi^2 \alpha^2 |\mathbfss{A}^{-T}\mathbf{k}|^2\big)^{3/2}},
\end{equation}
where $\mathbf{k}=(u,v)^T$ is measured in wavenumber units, $\beta=-0.7$ is the assumed spectral index for the synchrotron radiation emitted by the galaxy disc, $(l,m)$ and $\alpha$ are the source position and scalelength respectively, $S_{\lambda_\textrm{ref}}$ is the source flux at reference wavelength $\lambda_\textrm{ref}$. The ellipticity parameters $(e_1, e_2)$ are contained in the matrix~$\mathbfss{A}$ that linearly transforms the circular exponential profile to an ellipse:
\begin{equation}\label{linearTransf}
\mathbfss{A}  = \left( \begin{array}{cc} 1-e_{1} & -e_{2} \\ -e_{2} & 1+e_{1} \end{array} \right).
\end{equation}
We assume the following ellipticity definition:
\begin{equation}
\mathbf{e} = e_1 +\mathrm{i}e_2 = \frac{a-b}{a+b}\mathrm{e}^{2\mathrm{i}\theta},
\end{equation}
where $a$ and $b$ are the galaxy major and minor axes respectively, and $\theta$ is the galaxy orientation.

The likelihood is marginalised over non-interesting parameters such as flux, scalelength and position, adopting uniform priors for the flux and position, and a lognormal prior dependent on the flux for the scalelength (see Section~\ref{sec:ska}). This way we obtain a likelihood function of only the ellipticity parameters.
The galaxy ellipticity measurement is given by the likelihood mean point and 1D standard deviation (defined as the square root of the covariance matrix determinant), obtained after sampling the likelihood with an adaptive grid covering a neighbourhood around the maximum point.
 
In real observations the finite channel bandwidth and sampling time introduce smearing effects that are dependent on the source position. These effects may be approximated analytically and included in the visibilities model \citep{Smearing99,Smirnov11}. For example for frequency smearing, assuming a square bandpass filter in the expression of the smeared visibility presented in \cite{Smearing99}, we obtain:
\begin{equation}
\tilde V(u,v)= V(u,v) \textrm{sinc}[\pi(ul+vm)\Delta\nu/(\nu_0 c)],
\end{equation}
where uv coordinates are taken at the mid-channel frequency~$\nu_0$, $\Delta\nu$ is the channel bandwidth and $\text{sinc}(x)=\sin(x)/x$. 
Another option is to make the observation with very tiny frequency channels and sampling time intervals. \cite{SKA-ECP} proposed to use $\sim$30~kHz channel bandwidth and 0.5~s sampling time to make smearing tolerable, but meaning a huge number of uv points. In this case, raw visibilities may be averaged into a single uv grid without jeopardizing ellipticity measurements. In fact, observations from the same pair of antennae at different frequencies (resp. times) correspond to visibilities evaluated at different uv points along a radial (resp. tangential) direction, therefore these visibilities can be treated as the ones evaluated at uv points related to different baselines. Depending on the grid size, data volume and then computational time may be considerably reduced.

\subsection{Source extraction}
\label{sec:extraction}
We assume flux and source positions are provided. For example they may be estimated from a cleaned image of the same data that are analysed, or applying MC methods to the visibilities (e.g. using a multimodal nested sampling with a single source model as in \cite{FMH08}).
From this information we define an initial sky model where the visibilities of each source $s$ in the field of view are computed according to equation~(\ref{model}) with ellipticity $\mathbf{e}= \mathbf{0}$, i.e. circular source, and scalelength provided by the linear relation between the log of the median scalelength $\alpha_\mathrm{med}$ and flux density $S$ \citep{Rivi15}:
\begin{equation}
\ln{[\alpha_\mathrm{med}/\textrm{arcsec}]} = -0.93 +0.33\ln{[S/\mu \textrm{Jy}]}.
\label{scale-flux}
\end{equation}
The sky visibilities are obtained adding the model visibilities of each source in the beam: 
\begin{equation}
\label{sky}
V_\mathrm{sky}(u,v) = \sum_{s=1}^N V_s(u,v).
\end{equation}
Starting from this sky model, the source extraction and fitting procedure is performed according to decreasing flux order, i.e. decreasing SNR, as follows:  
\begin{itemize}
\item[1.] Given the position of the source $(l,m)$, remove the corresponding circular source model visibilities from the sky model and then take the difference between the data and the sky model, so that the visibilities of the current source (with a reduced contamination from the others) are isolated.
\item[2.] Apply \textit{faceting} \citep{CP92}: phase shift these visibilities in order to move the phase centre to the location of the source, by multiplying each visibility by the factor  
$\exp(-2\pi \mathrm{i} (u l + v m))$, and average them in a coarse grid (facet). This way we reduce the field of view to a small patch around the source, with the advantage of reducing the number of visibilities used for the model fitting and therefore accelerating the computation. On the other hand this procedure limits the maximum wavelength of the Fourier mode that can be measured because of the finite spacing of the facet uv points.
\item[3.] Use the source visibilities for measuring the corresponding source ellipticity as in Section~\ref{sec:fitting}.
\item[4.] Use the estimated ellipticity to improve model visibilities of the current source and remove them from the data.
\item[5.] Repeat from step~1 until all sources are fitted.
\end{itemize} 
Note that in this algorithm the sky model is improved after each source fitting by replacing circular sources with the elliptical source that has been fitted. Moreover, by ordering the source extraction by decreasing flux, the source fitting is performed with a better approximation of the sky model for sources at low SNR.

In the case of ''bad measurements'', the corresponding sources are not removed in the first instance from the data and sky model visibilities, but they are re-fitted at the end of the procedure, when the ellipticities of all the other sources are measured and a better sky model is obtained. These unreliable fits are recognised by a too small standard deviation of the ellipticity likelihood to be realistic. This seems related to errors in the likelihood computation, when the cross-correlation function is not sufficiently smooth to be marginalised over the source position, possibly due to PSF sidelobes or too much noise in the data. Bad measurements are given weight zero in the shear computation.

\section{SKA1 Simulations}
\label{sec:ska}
As in Paper~I, we simulate SKA-MID eight-hour observations of 1 square degree at declination $\delta = -30^\circ$ by using the SKA-MID\footnote{SKA-MID latitude is -30\degr 49\arcmin 48.00\arcsec~S. } Phase~1 antennae configuration  provided in \cite{SKA1a}. This integration time provides a complete circular coverage, i.e. without large gaps (because of the 3 telescope arms), and allow to reach a sensitivity of 10$\mu$Jy at 10$\sigma$. It would allow a targeted area of $800 \deg^2$ to be observed with such sensitivity in 10,000 hours in a forthcoming SKA1 radio weak lensing survey, sufficient for measuring cluster lensing. 
 
We choose the following conservative approximation of the frequency bandwidth: 950 - 1190 MHz, as proposed in \cite{BHCB16}. This seems to be the optimum observation frequency for a weak lensing survey with SKA1-MID, in case only 30\% of the full bandwidth of SKA Band~2 is usable (because of RFI problems, other surveys commensality, etc.). 
Visibilities are sampled every $\tau_\textrm{acc} = 60$~s  and we consider one large channel because smearing effects are not included on shorter time and bandwidth scales.
The observations are simulated by using equations~(\ref{model}) and (\ref{sky}) and adding an uncorrelated Gaussian noise whose variance is given by \cite{Sensitivity99}:
 \begin{equation}
 \label{final_variance}
 \sigma^2 = \frac{\textrm{SEFD}^2}{2\eta^2\Delta\nu \tau_\textrm{acc}},
 \end{equation}
where $\Delta\nu$ is the frequency channel bandwidth, SEFD = 400~Jy is the system equivalent flux density for SKA-MID dishes and $\eta=0.9$ is the system efficiency \citep{SKA1b}. For simplicity we assume that the SKA1-MID core is composed of only SKA dish antennae even if part of it actually contains 64 MeerKAT dishes.

SF galaxy populations are generated according to distributions estimated from archival data of VLA radio surveys at 1.4~GHz. As described in \cite{Rivi15},
we estimated distributions of the flux $S$ and the scalelength~$\alpha$ of sources modelled by an exponential profile by the analysis of faint sources (order of tens $\mu$Jy) catalogue of the SWIRE field survey \citep{OM2008}. This is the radio catalogue with the largest number of SF galaxies, being related to the deepest survey so far in terms of the radio source density (source flux cut $\sim$ 15 $\mu$Jy), and it contains source size measurements from the imaged data. The flux distribution is fitted by a power law: $p(S) \propto S^{-1.34}$.  
The scalelength is obtained by the relation with the measured full width at half maximum\footnote{The FWHM is derived from the Gaussian model fit of the source after PSF deconvolution.}: FWHM =~$2\alpha\ln(2)$, and its distribution is fitted dependently on source flux by a lognormal function with  mean $\mu = \ln(\alpha_\mathrm{med})$ and variance $\sigma \sim 0.3$, where $\alpha_\mathrm{med}$ is given by eq.~(\ref{scale-flux}). The variance value is suitably chosen in the middle of a range well representing scalelength distributions for different flux values.
The modulus $e$ of the intrinsic ellipticities is generated according to a function proposed by \citet{Miller13}:
\begin{equation}
P(e) = \frac{Ne\left(1-\exp\left[ \frac{e-e_\textrm{max}}{c}\right]\right)}{(1+e)(e^2+e_0^2)^{1/2}}.
\end{equation}
The parameter values, $c =  0.2298$ and $e_0 = 0.0732$ are obtained from fitting the VLA-COSMOS field data. Although this survey is less deep than the SWIRE, but still detecting $\mu$Jy sources (flux cut $\sim$ 75 $\mu$Jy), we rely on a recent re-analysis of the L-band radio visibility data where the level of systematics in the measurement of the galaxy position angle is significantly reduced \citep{Tunbridge16}. In fact the previous analysis \citep{COSMOS07} as well as the one of the VLA-SWIRE were mainly focused on faint source counts.
The normalisation factor is $N =  2.595$.

We generate galaxy populations with flux densities ranging between  $10\mu$Jy and $200\mu$Jy. According to our flux distribution we obtain a source number density of 2.7~gal/arcmin$^2$. To be consistent, we adopt this source number density in our simulation, although more accurate modelling from recent radio continuum surveys suggest that a higher source number density should be detected at such a flux density cut in real observations \citep{Condon12, Mancuso17}. This is the expected source number density for the proposed 2-yr SKA1 radio weak lensing survey covering 5000 $\deg^2$ \citep{Brown15}.

\section{Galaxy shape measurement}
\label{sec:shape}
In this section, first we select the facet size to be used in the source extraction by simulating visibilities of individual sources. Then we simulate populations of galaxies located simultaneously in the field of view in order to  show the efficacy of our source extraction algorithm as a function of the source number density. 

\subsection{Facet size}
\label{faceting}

 \begin{figure*}
\includegraphics[scale=0.58]{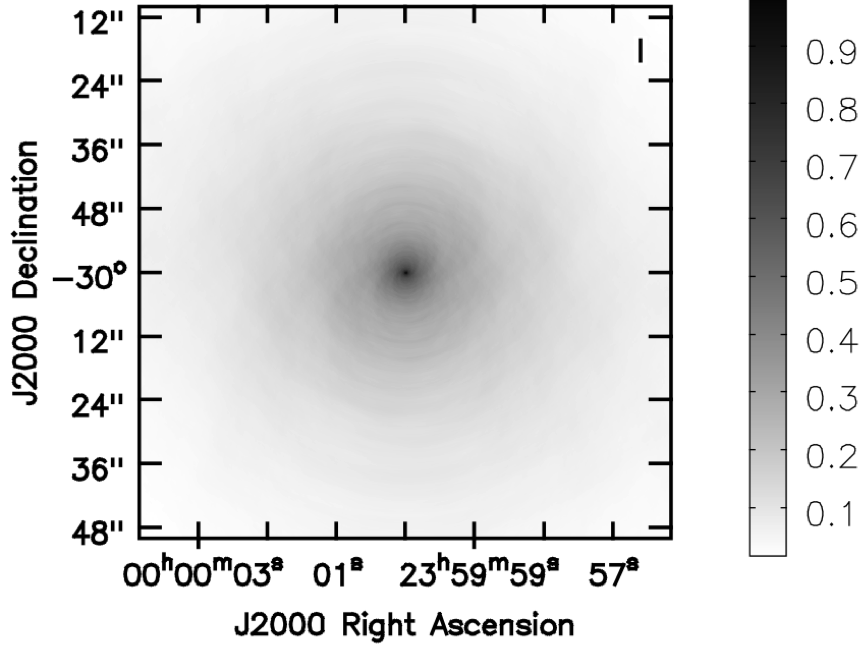}
\includegraphics[scale=0.59]{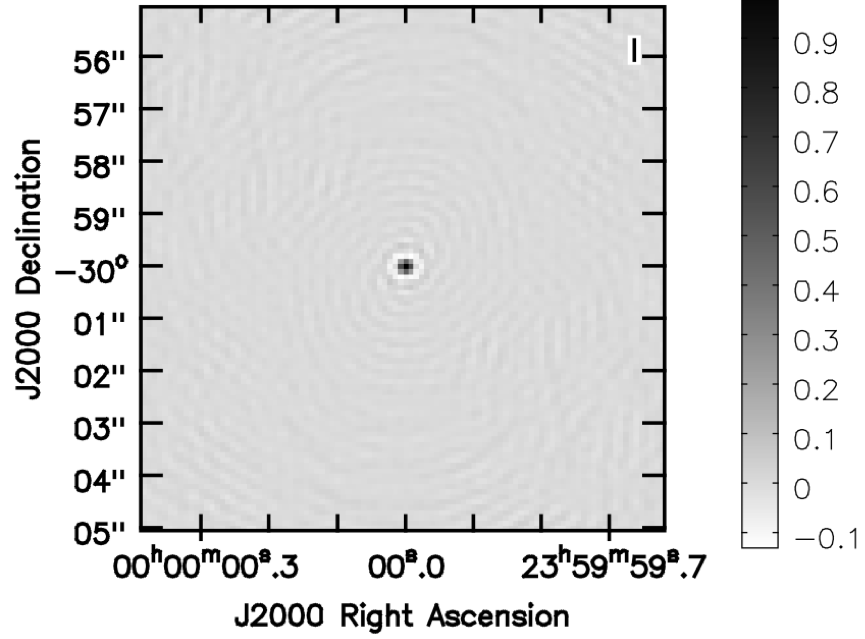}
\caption{Image of the PSF corresponding to our SKA1 uv coverage \textit{Left}: Natural weighting. The PSF has a broad, low-level plateau because uv points have short spacings close to zero, as they tend to spend more time per unit area near the uv origin. \textit{Right}: Uniform weighting.}  
\label{fig:PSF}
\end{figure*}

\begin{table}
\begin{tabular}{l c c c }
\hline
 S [$\mu$Jy] & n. cells  & \multicolumn{2}{|c|}{best-fit slope}  \\
 & & $e_1$ & $e_2$ \\
\hline
 200-150 & 600  & $0.9774 \pm 0.0025 $ & $0.9675 \pm 0.0025$ \\
 150-100 & 550  & $0.9795 \pm 0.0030 $ & $0.9664 \pm 0.0030$ \\
 100-80  & 500 & $0.9797 \pm 0.0032$ & $0.9660 \pm 0.0032$ \\
  80-60  & 460 & $0.9774 \pm 0.0029$ & $0.9614 \pm 0.0029$ \\
  60-40  & 420 & $ 0.9760 \pm 0.0031$ & $0.9631 \pm 0.0031$ \\
  40-20  & 350  & $0.9756 \pm 0.0028 $ & $0.9557 \pm 0.0029$\\
  20-10 & 280 & $0.9765 \pm 0.0030$ & $0.9510 \pm 0.0030$\\
\hline
\end{tabular}
\caption{Facet sizes dependent on source flux range and corresponding best-fit slopes for 1000 sources.}
\label{tab:facet_size}
\end{table}

\begin{figure*}
\includegraphics[scale=0.41]{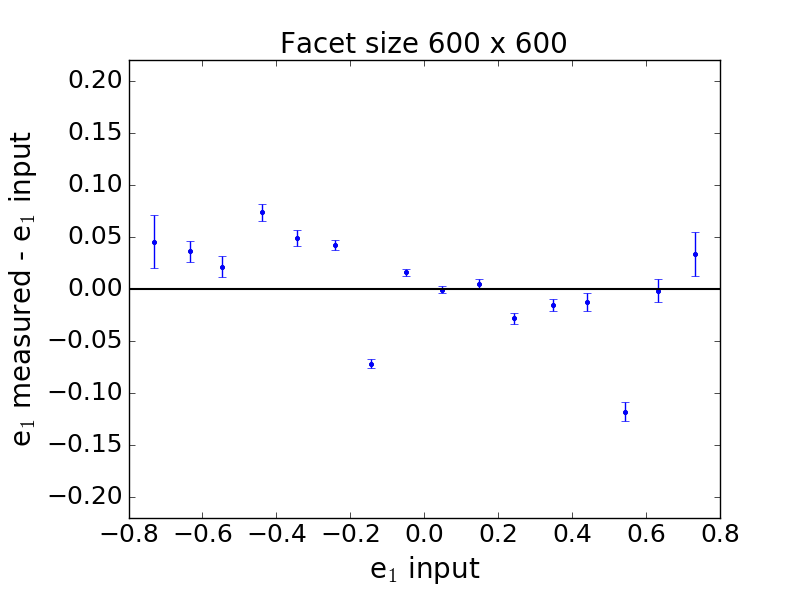}   
\includegraphics[scale=0.415]{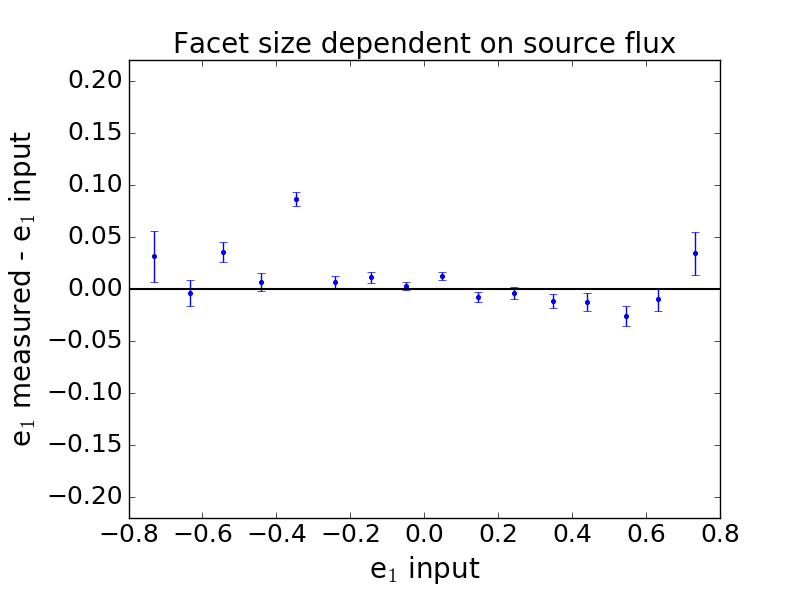} 
\caption{Binned measurements (minus true values) of the ellipticity first component of 1000 individual galaxies randomly located in the field of view and with flux ranging between 10~$\mu$Jy and 200~$\mu$Jy. \textit{Left}: facet size $600 \times 600$ for all sources. \textit{Right}: facet size dependent on source flux.}
\label{fig:single-bestfit}
\end{figure*}

The facet size is affected by the weighting scheme \citep{Weighting99} adopted in the gridding phase. For example, \textit{natural} weighting optimises the sensitivity for detecting weak sources by emphasising the data from short baselines. In this case, a relatively large facet size is expected even for covering a single galaxy because a large contribution to the signal is from long wavelength modes which must be adequately sampled by small facet cells. In effect, the source in the image domain turns to be convolved with a large natural-weighted PSF with a broad low-level plateau (see left panel of Fig.~\ref{fig:PSF}). \textit{Uniform} weighting will require instead a much smaller facet size because it emphasises data from long baselines, where most of the source shape signal is contained. This is reflected by the small uniform-weighted PSF (see right panel of Fig.~\ref{fig:PSF}).
On the other side, the weighting scheme used in the faceting procedure shouldn't affect the model fitting, provided that the measurement uncertainties are propagated correctly in the likelihood computation (see equation (21) of Paper~I for the natural case) and model visibilities are consistent with the observed data. This may not be true for measurement methods in the image domain, as shown in \cite{Tunbridge16}.
 
Since we are interested in the detection of faint sources for radio weak lensing, we adopt a natural weighting scheme.   
To minimise the number of sources falling in the same facet, we define a facet size dependent on source flux, as it is related to the size of the source.  We split the flux total range of the simulated population, i.e. 10-200 $\mu$Jy, in 7 bins as shown in Table~\ref{tab:facet_size}. Facet uv point coordinates have to be re-computed only once per bin as the model fitting is performed according to source flux order. We chose larger bins at large fluxes because the sizes of such sources increase more gradually with the flux compared to the ones with low flux (see equation~(\ref{scale-flux})). 
To estimate the facet size for each bin, we simulate raw visibilities of a single galaxy in the primary beam in order to avoid any source contamination effects, and vary the noise added to the visibilities (to have SNR~$\ge 100$) in order to see the effect of the source size only. We measure the galaxy ellipticity after averaging visibilities in the facet.
The best-fit slope\footnote{Consistently with Paper~I, we refer to the ellipticity best-fit slope instead of the multiplicative bias when measuring galaxy shapes. This terminology is used to clearly distinguish it from the shear multiplicative bias, which is obtained from the best-fit of shear measurements (each being the weighted average of galaxy ellipticities).} for the ellipticity measurements of 1000 galaxies is computed for different facet sizes and source flux ranging between the flux bin bounds. We select the facet size when a fixed best-fit slope threshold of about 0.97 is reached, as listed in Table~\ref{tab:facet_size}.
 
Note that the fitting for the first ellipticity component is better than the second one because of a slightly anisotropy of the PSF as discussed in Section~\ref{sec:PSF}.  
The selected facet sizes are consistent with the relation between the uv grid cell size $\Delta u$ (in units of wavelengths) and the related field of view (in radians): $\psi = 1/\Delta u$ \citep{Weighting99}.
We also note that for small sources (low flux) it is actually better to use smaller facets  even in the case of a single source in the field of view. This is shown in Fig.~\ref{fig:single-bestfit} where the difference between binned measurements and true values of the first ellipticity component of 1000 galaxies, with realistic flux distribution in the range 10-200~$\mu$Jy, are plotted both for the case where the facet size is constant and equal to 600 (left panel) and when the facet has a variable size dependent on the source flux (right panel).  Similar plots are obtained for the second component.  In the latter case we obtained 25 bad measurements (see Section~\ref{sec:extraction}) and the best-fit slopes of the two ellipticity components are $0.9552 \pm 0.0057$ and $0.9426 \pm 0.0061$ respectively, whereas for the case of $600 \times 600$ facet the best-fit slopes for the same galaxy population and noise are $0.9306 \pm 0.0054$ and $0.9135 \pm 0.0056$ and the number of bad measurements is 3 times larger.

These results are due to the fact that we do not model exactly the primary beam because the model visibilities are directly sampled on the uv facet points. This means that in the image domain the sidelobes of the source model are not suppressed by any apodisation, whereas the gridding of the original uv coverage causes the full image to be apodised by a broad 2D sinc function which has the effect in the data of suppressing background sources that are a long way from the phase centre and the distant sidelobes from the primary source. The grid sampling causes the resulting image domain facet to become a small, but aliased version of the apodised image. The aliased model is an incorrect description of the apodised and aliased data, and the discrepancy will get worse for smaller facets and at large distances from the source. Fig.~\ref{fig:single-bestfit} shows that a suitable facet size dependent on the source flux/size may be a trade-off between these two effects.
We could improve the model by applying the same gridding operations as in the data (sampling on the original uv coverage and then averaging in the facet), but this will add a large amount of computational time. Our results show that the adopted model approximation is acceptable, provided that the facet sizes are sufficiently large to not affect the significant sidelobes in the image domain. Otherwise we expect the discrepancy between data and model to become severe and the biases may become less robust and hence less calibratable.

\subsection{Dependence on source number density}
\label{sec:density}

\begin{figure}
\includegraphics[scale=0.45]{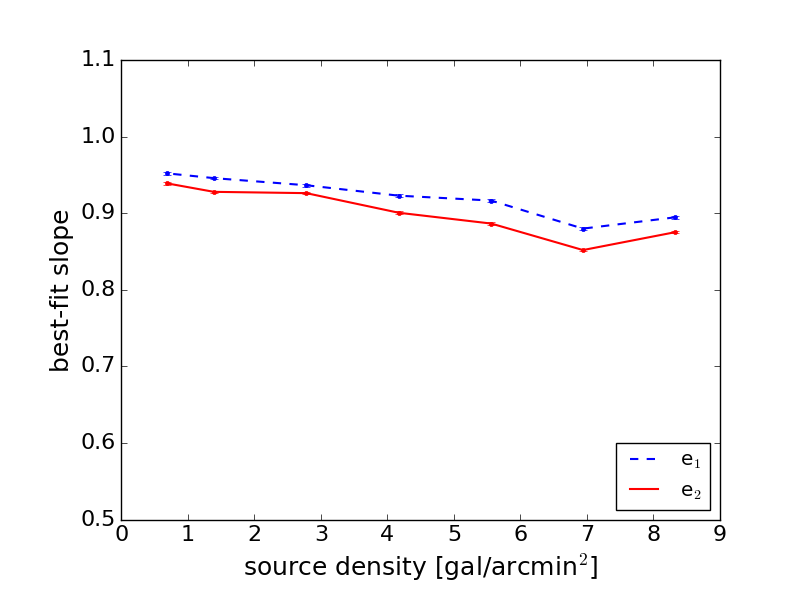}
\caption{Best-fit line slope for both galaxy ellipticity components as a function of the source number density.}
\label{fig:nSources-shape}
\end{figure}

We estimate the efficacy of the source extraction method by measuring the slope of the best-fit line of $10^4$ galaxy ellipticity measurements as a function of the source number density. For each measurement we simulate sources located simultaneously in the field of view according to a uniform distribution. Results are plotted in Fig.~\ref{fig:nSources-shape} and show reliable fits, independent of the source density up to 2.8 gal/arcmin$^2$.
In this case the best-fit slopes of the correlation for each ellipticity component are $0.9365 \pm 0.0017$ and $0.9262 \pm 0.0017$ respectively and the number of bad measurements is about 1\%. At higher densities galaxy ellipticity measurement starts to deteriorate, as residuals of nearby galaxies affect the model fitting, but may still be good enough for shear measurement because of the improved statistics (as shown in Section~\ref{sec:shear}). 
Shape measurements of galaxies may be improved by a joint fitting within facets by applying the Hamiltonian Monte Carlo technique \citep{Neal11}. \textit{RadioLens}fit results used as starting points should reduce the burn-in phase and accelerate convergence. Since the number of sources in the facet will be relatively small this approach becomes more feasible and preliminary results with this method show a better accuracy in the galaxy ellipticity fitting, although requiring a large computational time (Rivi et al., in preparation). 

Using a single channel, the serial version of \textit{RadioLens}fit running on an Intel Xeon E5-2650 takes on average about 10~sec/gal computing time for the model fitting.
As discussed in Paper~I, the shared memory parallelisation with OpenMP allows to exploit all the computational resources when the amount of memory for the source model fitting requires the usage of the full CPU. Its implementation has been optimised by distributing to each thread the likelihood computation and marginalisation over the position parameters for different scalelength values of the model. It doesn't scale linearly with the number of threads because the likelihood mariginalisation over the scalelength parameter is not parallel and there is an overhead for the creation and destruction of OpenMP threads at each iteration of the likelihood maximisation and sampling. This version running on all the eight cores of the CPU takes on average about 2.4~sec/gal.

\section{Shear} 
\label{sec:shear}
Following Paper~I, for shear measurement we simulate galaxy populations as described in Section~\ref{sec:ska} in a field of view of 1~$\deg^2$. We generate populations free of shape noise \citep{Nakajima07, Massey07}: for each ellipticity modulus, 10 equally-spaced galaxy orientations are generated so that the corresponding ellipticity values are distributed uniformly on a circle, and galaxies whose ellipticity is on the same ring are given the same size and flux. 
We generate sources randomly located according to a uniform distribution.
All measurements are performed with facet size dependent on the source flux as defined in Table~1.

\begin{figure}
\includegraphics[scale=0.45]{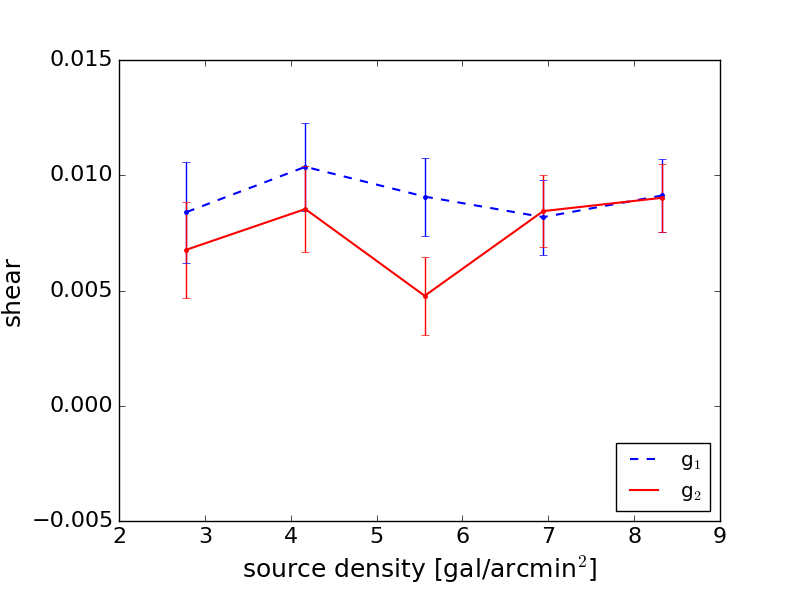}
\caption{Shear components estimated from a galaxy population in 1 $\deg^2$ as a function of the source number density for input $\mathbf{g} = \mathbf{0}$.}
\label{fig:nSources-shear}
\end{figure}

We measure the reduced shear $\mathbf{g}=g_1 + \mathrm{i}g_2$ as the weighted average of the galaxies' ellipticities using weights that approximate the inverse-variance of each ellipticity measurement. Error bars are given by the standard deviation of the shear values estimated  from 1000 bootstrap resamples. Fig.~\ref{fig:nSources-shear} shows shear measurements as a function of the source number density up to 8 gal/armin$^2$, when no input shear is applied, i.e. $g_1=g_2=0$. At high densities the larger number of sources compensates the less accurate galaxy shape fitting (Fig.~\ref{fig:nSources-shape}), still producing shear values consistent with the results obtained at the SKA1 source density corresponding to a population of about $10^4$ galaxies.

\begin{figure*}
\includegraphics[scale=0.462]{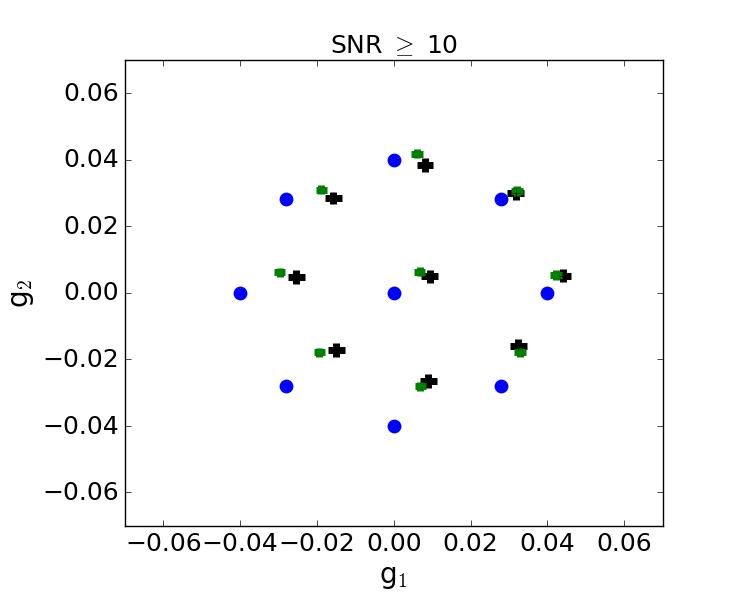}
\includegraphics[scale=0.466]{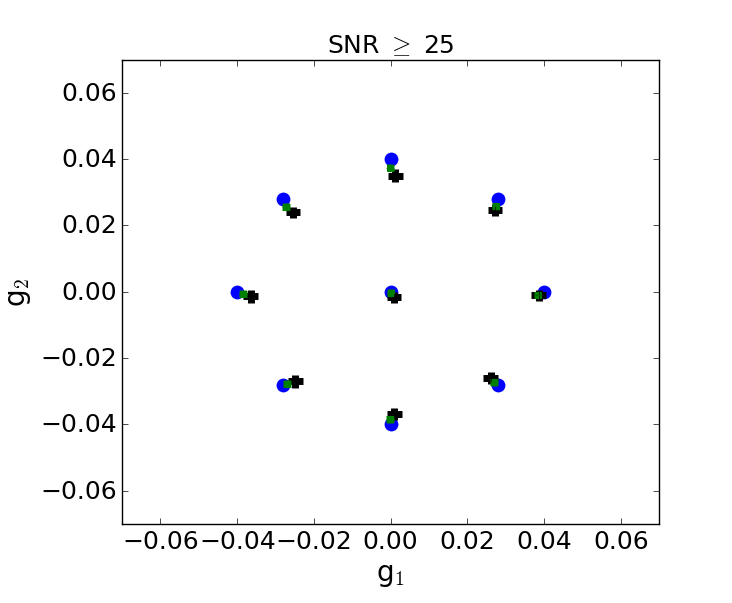}
\caption{Comparison of shear measurements: input values are blue points, measured values from single sources at the phase centre are green crosses, measured values from sources simultaneously in the f.o.v are black crosses.} 
\label{fig:shear}
\end{figure*}

For this population we measure the shear also for input reduced shear values with amplitude $g=0.04$ and eight different orientations.   
The input shear $\mathbf{g}$ action on the intrinsic galaxy ellipticity $\mathbf{e}^s$ is simulated following \cite{SS97}:
\begin{equation}
\mathbf{e} = \frac{\mathbf{e}^s + \mathbf{g}}{1+\mathbf{g}^*\mathbf{e}^s},
\end{equation} 
where $\mathbf{g}^*$ is the shear complex conjugate.
We compare results with the optimal case where each galaxy is at the phase centre and the only one contained in the field of view, considering the same galaxy population. 
Results are plotted in Fig.~\ref{fig:shear}, both for SNR $\ge 10$ and SNR $\ge 25$, where measurements from individual source visibilities are green crosses and the ones from the same population but with all sources simultaneously in the primary beam are black crosses. Clearly error bars (cross arms) are larger at SNR~$\ge 10$ as the galaxy population is dominated by lower flux sources.

\begin{table*}
\begin{tabular}{l l c c c c }
\hline
& & $m_1$ & $c_1$ & $m_2$ & $c_2$  \\
\hline
SNR $\ge 10$ & single source &  $-0.0904 \pm 0.0186$ & $0.00655 \pm 0.00050$ & $ -0.1297 \pm 0.0171$ & $ 0.00632 \pm 0.00046$ \\
 & multiple sources & $ -0.1428 \pm 0.0274$ & $ 0.00872 \pm 0.00073$ & $-0.1864\pm 0.0256$ & $0.00578 \pm 0.00067$ \\
\hline
SNR $\ge 25$ & single source & $-0.0352 \pm 0.0127$ & $ -0.00006 \pm 0.00034$ & $-0.0506\pm 0.0132$ & $-0.00070 \pm 0.00035$ \\
& multiple sources &  $-0.0677 \pm 0.0242$ & $ 0.00089 \pm 0.00064 $ & $ -0.0992 \pm 0.0230$ & $ -0.00112 \pm 0.00060$ \\
\hline 
\end{tabular}
\label{tab:shear-bias}
\caption{Shear bias components estimated from a realistic population of $\sim 10^4$ galaxies randomly distributed in 1 $\deg^2$, corresponding to a source number density of 2.8 gal/arcmin$^2$.}
\end{table*}

The measured shear bias, defined as
\begin{equation}
g_i^m - g_i = m_i g_i + c_i, \qquad i=1,2,
\end{equation}
is shown in Table~2. 
At SNR $\ge 10$, the multiplicative biases~$m_i$ for the two shear components are respectively 1.6 and 1.4 times the ideal case of a single source in the field of view, while additive bias~$c_i$ are almost consistent. Selecting galaxies with SNR $\ge 25$ the population reduces to 5810 sources (i.e. 1.6~gal/arcmin$^2$). This should affect the bias uncertainty only,
as the bias on the shear measurement should not depend on the number of sources. At this regime the $m$ values reduce by a factor two instead of three as in the single source case. 
This is due to the source signal contamination by residuals of nearby galaxies, which is a new contribution to the shear bias that seems to have an effect on the multiplicative terms only. It may be mitigated by refining ellipticity measurements by joint fitting within larger facets, as explained in Section~\ref{sec:density}.  
Note that ''neighbour bias'' affects optical lensing measurements also (see \citealt{Miller13}, \citealt{Jarvis16}, \citealt{Mandelbaum17}, \citealt{Samuroff17} and \citealt{ Zuntz17}). 

As discussed in Paper~I, the noise bias values exceed SKA1 survey requirements\footnote{For a 2-yr SKA1-MID weak lensing survey over 5000~$\deg^2$ and $z_\mathrm{med} = 1.0$ the requirements for cosmological parameters measurements to be dominated by statistical rather than systematic errors are: multiplicative bias $m < 0.0067$, additive bias $c < 0.00082$ \citep{Brown15}. They are derived using the rules provided in \cite{AR2008}.} except for the additive component at SNR $\ge 25$. However they are comparable to the ones obtained from optical surveys using \textit{lens}fit \citep{KiDS17} and {\sc im3shape} \citep{Zuntz17}, where a shear calibration correction reduced the multiplicative bias to well below the percent level. Standard approaches in the optical domain derive such calibration by inferring the bias from simulated data matching the observations \citep{Bruderer16} or parametrising the bias as a function of the observed galaxy properties \citep{Kuijken15, Jarvis16}.
Recently a self-calibration approach, implemented in the \textit{Metacalibration} method \citep{Metacal17, SH2017} and used in the analysis of the Dark Energy Survey\footnote{https://www.darkenergysurvey.org} (DES) \citep{Zuntz17}, proved to be the most efficient, being able to recover the input shear in realistic simulations to better than a part in a thousand. It also isotropises the PSF to remove any additive bias.
The key idea of the method is to compute the shear estimator response for a shape measurement directly from observed data perturbed with a small known shear. This way all the features present in real data are already included, which are instead extremely difficult to model accurately in external simulations, and it can be applied to any shear measurement method based on averages of galaxy shapes. A similar approach may then be implemented quite easily in the interferometer data analysis, with the advantage that for the additive bias at radio wavelengths we know the PSF much better than at optical wavelengths and we can make the PSF isotropic directly by weighting the uv-plane (as discussed in Section~\ref{sec:PSF}).

\section{Additive bias dependence on the image-plane PSF shape}
\label{sec:PSF}

It is well known from weak lensing optical surveys that shear additive bias is dependent on the PSF shape \citep{Miller13}. For radio interferometers the PSF is deterministically defined by the uv coverage of the telescope (i.e. antennae locations and pointing direction).  For example, if we increase the antenna pointing declination to a larger zenith distance the PSF shape becomes compressed along the y-axis. 

\begin{figure}
\includegraphics[scale=0.45]{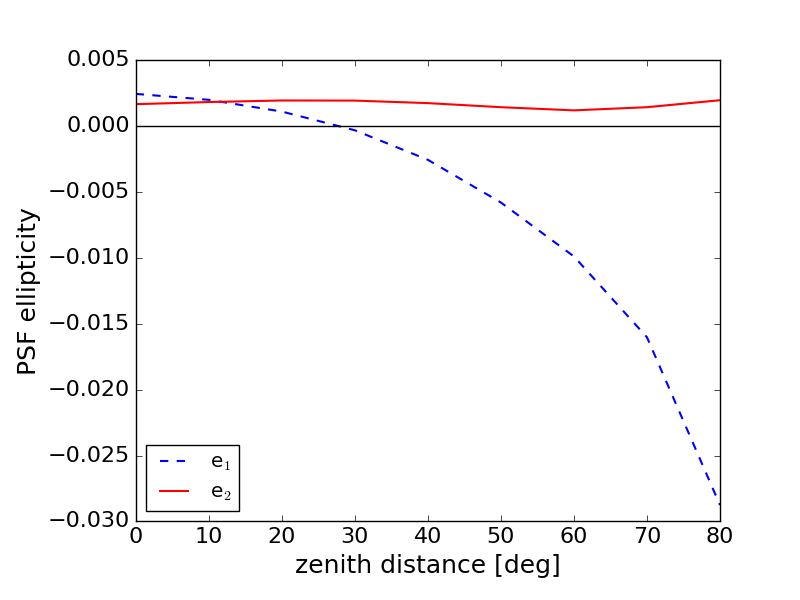}
\caption{PSF ellipticity components versus the zenith distance at the starting of the observation.}
\label{fig:PSF-dec} 
\end{figure} 

We measure the additive noise bias for different pointing declinations at various SNR values. Starting from our uv coverage (corresponding to the same declination as the observatory latitude), we simulate the effect on the uv points when the phase centre declination is increased by an angle $\phi$.
A plot of the image-plane PSF ellipticity components as functions of the zenith distance is given in Fig.~\ref{fig:PSF-dec}.
The R-squared size of the PSF slightly increases from  14.83~arcsec$^2$ to 15.54~arcsec$^2$. These values are computed from the quadrupole moments of the image domain as follows \citep{Schneider2006}:
\begin{align}
& \mathbf{e} = \frac{Q_{xx}-Q_{yy}+2\mathrm{i}Q_{xy}}{Q_{xx}+Q_{yy}+2(Q_{xx}Q_{yy}-Q^2_{xy})^{\frac12}}, \\
& R = Q_{xx}+Q_{yy}.
\end{align}

We simulate individual source visibilities, to avoid nearby source contamination effects, and assume a constant maximum facet size $1000 \times 1000$ to ensure that the galaxy convolved with the PSF is contained in the facet even when the PSF becomes highly anisotropic. 

\begin{figure*}
\includegraphics[scale=0.405]{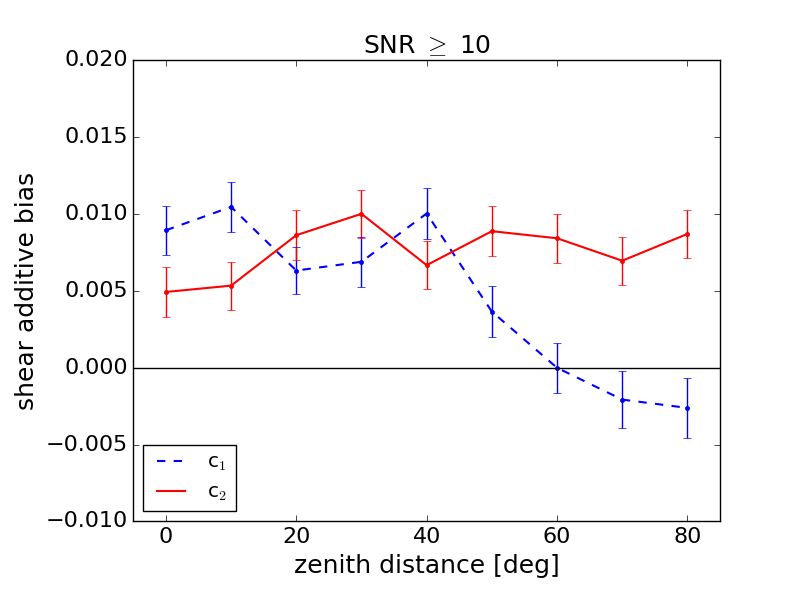}
\includegraphics[scale=0.41]{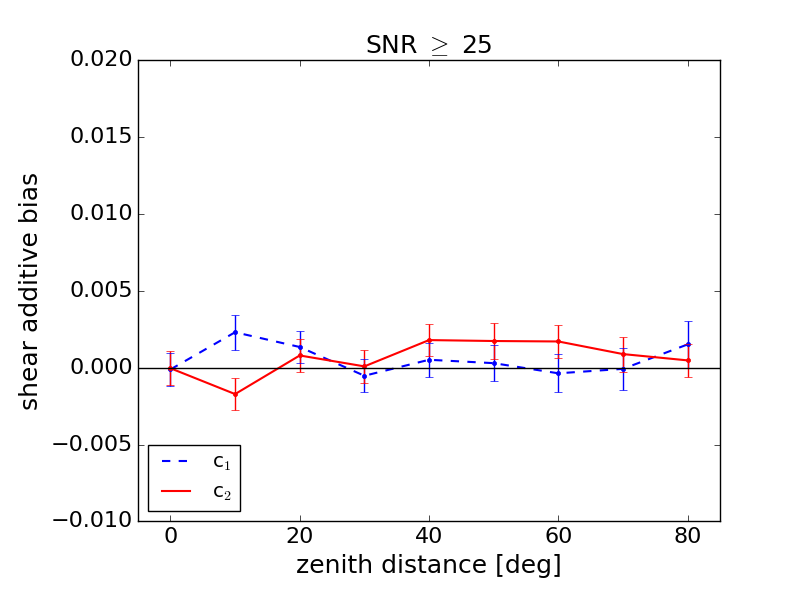}
\caption{Shear additive bias as a function of the zenith distance at different lower signal-to-noise. Facet size fixed at $1000 \times 1000$.} 
 \label{fig:plot-dec} 
\end{figure*}

\begin{figure*}
\includegraphics[scale=0.41]{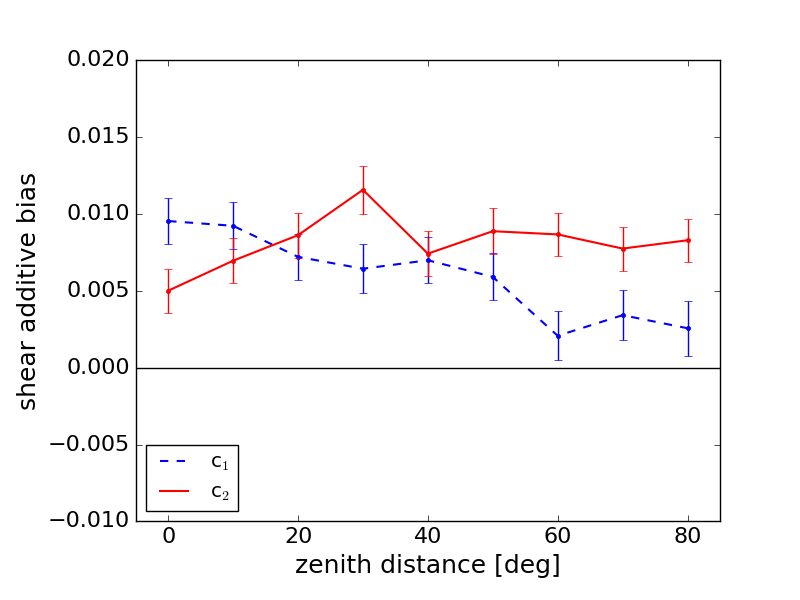}
\includegraphics[scale=0.41]{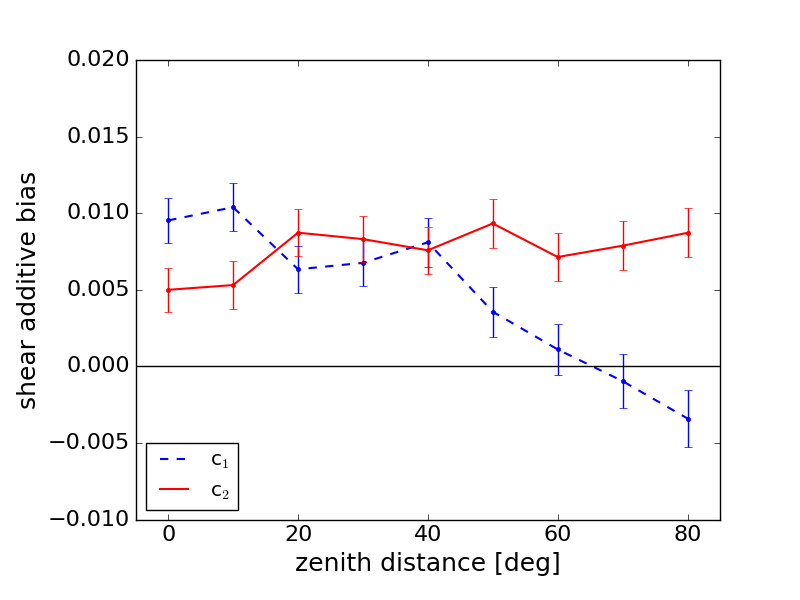}
\caption{Shear additive bias at SNR $\ge 10$ as a function of the zenith distance. \textit{Left}: facet size dependent on source flux according to Table~\ref{tab:facet_size} for all declinations. \textit{Right}: facet size increased with source declination produce similar results as for a constant large facet size.}
\label{fig:SNR10-dec}
\end{figure*}

We observe that the additive bias is dependent on source size. In fact measurements at the same SNR obtained by lowering  the noise instead of increasing the source flux cut, produce larger bias values, meaning that the additive bias worsens when source sizes decreases. This is consistent with the analysis presented in \cite{Massey13}.
Noise bias causes a correlation between measured shear and PSF ellipticity even when we correct for the PSF in the model fitting.  This becomes noticeable at low SNR, where the first ellipticity component increases significantly towards larger negative values, as the PSF becomes more compressed along the y direction (see left panel of Fig.~\ref{fig:plot-dec}). At large SNR the additive bias almost disappears independently of the PSF shape (see right panel of Fig.~\ref{fig:plot-dec}).  This is in good agreement with what we should expect. 
Because of our long integration time the PSF is not isotropic even when the declination equals the observatory latitude, as in the baseline simulation, besides the fact that the distribution of the SKA-MID baselines on the ground are not circularly symmetric. Therefore a large additive bias is still measured at small zenith distances. 
The PSF anisotropy may be reduced by combining snapshots obtained over a range of hour angles, however this may not be sufficient to reach an additive bias acceptable for weak lensing surveys. To further reduce the noise bias at low SNR we also need to weight the uv plane to ensure that the PSF is more isotropic. A standard technique in radio imaging to improve PSF shape is to use tapering functions~\citep{Weighting99} to define uv points weights, although a more specific weighting scheme may be required.  Moreover, as for the multiplicative bias, we can calibrate the additive shear bias with simulations. This is more feasible than in optical surveys because the PSF is deterministic
at radio wavelengths.  However, any such calibration would be strongly dependent on distributions of source properties, so isotropising the PSF is a much better option. 

Note that when using variable size faceting, the facet size must be dependent not only on the source size but also on the PSF shape. In fact as the PSF becomes anisotropic the facet size may become too small relative to the size of the source convolved with the PSF, modifying the effective shape of the source. For example the left panel of Fig.~\ref{fig:SNR10-dec} shows what happens at SNR~$\ge 10$ when we maintain the same flux dependent facet size (Table~\ref{tab:facet_size}) for all pointing declinations: 
at large zenith distances the shortest baselines occupy smaller uv frequencies and therefore can measure wavelengths longer than the limit imposed by the small facet size used to extract the majority of the galaxy population. 
If we increase the facet size according to the source declination, e.g. 50 cells per side every $10^\circ$ declination increment, we obtain consistent results with the case of one single large facet (see right panel of Fig.~\ref{fig:SNR10-dec}).

\section{Conclusions}

We have extended the presentation of the \textit{RadioLens}fit method, introduced in Paper~I for the simple case of individual galaxies located at the phase centre, to the real case where many galaxies are randomly located in the field of view. This has been done by isolating the visibilities of each source and shifting the phase centre to the source position so that a coarse grid can still be used to reduce nearby galaxy residuals contamination and accelerate ellipticity measurement computation. Source extraction has been performed by removing apart from the data the simulated visibilities of the sky model, but the source of interest, given the positions and fluxes of all sources in the field of view from the image, and down-weighting what remains of nearby source-contamination by averaging visibilities in a coarse grid (facet).  For gridding we adopted a natural weighting to maximise sensitivity and estimated the smallest facet size dependent on source flux thresholds in order to minimise the number of nearby galaxies included in the facet. 

We tested the source extraction procedure, simulating visibilities of SF galaxy populations observed by SKA1-MID in the first 30 per cent of frequency Band~2. We adopted flux and scalelength parameters distributions estimated from the VLA SWIRE catalogue and used the \textit{lens}fit ellipticity prior with coefficients fitted from a new version of the VLA COSMOS catalogue optimised on shape measurements.

We showed the efficacy of our source extraction algorithm as a function of the source number density, obtaining a reliable galaxy ellipticity fitting for the density expected from the current proposal of the SKA1 radio weak lensing survey.  Shear measurements from eight-hour observation of one square degree show that the bias due to the extraction procedure mainly affects the multiplicative bias as no significant change has been observed for the additive bias when comparing with the bias obtained for the ideal case of a single source at the phase centre at a time. This bias may be mitigated by a second step in the galaxy ellipticity measurement, where a joint fitting within the facets is performed with HMC, starting from the values obtained with \textit{RadioLens}fit. However multiplicative noise bias calibration is also required as for the optical domain.

We finally observed that because of our uv coverage the PSF is slightly anisotropic even if pointing close to the zenith, therefore we obtain a large additive bias (on average about $0.0068 \pm 0.0006$ at SNR $\ge 10$). Although a suitable choice of the integration time (split and distributed along a longer period of time) may reduce the PSF anisotropy,  a uv weighting scheme may still be required to satisfy weak lensing requirements. It should be optimised to avoid any significant reduction of the signal-to-noise.

\section*{Acknowledgements}
We thank Tom Kitching for useful discussions and Ben Tunbridge for providing the distribution parameters of our ellipticity prior, obtained by fitting VLA-COSMOS data as for the prior function presented in his paper. We also thank Sphesihle Makhathini for the support in the simulation of the SKA1-MID uv coverages.

MR acknowledges the support of the Science and Technology Facilities Council  (STFC) via an SKA grant. LM acknowledges STFC grant ST/N000919/1.

\bibliographystyle{mn2e_trunc8}
\bibliography{master}

\end{document}